\documentclass[useAMS,usenatbib,fleqn]{mn2e_arxiv}
\usepackage{graphicx,graphics,amsmath,amssymb}
\usepackage{url}
\usepackage{mathtools}

\DeclarePairedDelimiter{\erfcfences}{[}{]}
\newcommand{\erfc}{\operatorname{erfc}\erfcfences}

\title[The Galactic centre pulsar population]
{The Galactic centre pulsar population}

\author[Chennamangalam and Lorimer]
{Jayanth Chennamangalam$^{1}$\thanks{Email: jchennam@mix.wvu.edu} and D.~R.~Lorimer$^{1,2}$\thanks{Email: duncan.lorimer@mail.wvu.edu}\\
$^1$ Department of Physics and Astronomy, West Virginia University, PO~Box~6315, Morgantown,
WV~26506, USA\\
$^2$ National Radio Astronomy Observatory, PO~Box~2, Green Bank, WV~24944, USA
}

\begin{document}
\maketitle
\begin{abstract}
The recent discovery of a magnetar in the Galactic centre region has allowed
Spitler~et~al. to characterize the interstellar scattering in that direction.
They find that the temporal broadening of the pulse profile of the magnetar is
substantially less than that predicted by models of the electron density of
that region. This raises the question of what the plausible limits for the
number of potentially observable pulsars -- i.e., the number of pulsars beaming
towards the Earth -- in the Galactic centre are. In this paper, using
reasonable assumptions -- namely, (i) the luminosity function of pulsars in the
Galactic centre region is the same as that in the field, (ii) the region has
had a constant pulsar formation rate, (iii) the spin and luminosity evolution
of magnetars and pulsars are similar, and (iv) the scattering in the direction
of the Galactic centre magnetar is representative of the entire inner parsec --
we show that the potentially observable population of pulsars in the inner
parsec has a conservative upper limit of $\sim$ 200, and that it is premature
to conclude that the number of pulsars in this region is small. We also show
that the observational results so far are consistent with this number and make
predictions for future radio pulsar surveys of the Galactic centre.
\end{abstract}


\section{Introduction}

Discovering radio pulsars in the Galactic centre (GC) has been a long-sought goal, due
to the promise it bears in probing the gravitational field of the massive black hole in
the region \citep[e.g.,][]{pfa04, liu12}, and in deciphering the nature of the
interstellar medium in its vicinity. Despite several radio surveys
\citep[e.g.,][]{kra00,joh06,den09,den10,mac10,bat11,eat13a,sie13}, no pulsars were found.
So far, the only pulsar in the inner parsec of the GC was found to be a $\sim$~3.76~s magnetar,
which was discovered following an X-ray flaring episode in April
2013 \citep{ken13}, and a subsequent periodicity search \citep{mor13}. This was
later confirmed in the radio as PSR J1745$-$2900 \citep{eat13b,but13}. \cite{ken13},
\cite{mor13} and \cite{rea13} analysed the X-ray absorption of this source, and found that
it is consistent with being at a similar distance as Sgr A*. \cite{rea13} localised the
magnetar to an angular distance of $\sim 2.4''$ from Sgr A*. At the GC distance of 8.25 kpc
\citep{gen10}, this corresponds to a minimum distance of $\sim$ 0.1 pc. Interstellar
scattering was thought to be a major problem in detecting pulsars in the GC, as models of
electron density dictate a scattering timescale of at least $6.3~\nu_{\rm GHz}^{-4}$ s, but
potentially up to 200 times larger \citep{cor97}, indicating that observations at higher
frequencies are more favourable, with \cite{mac10} suggesting optimal frequencies in the
10--16 GHz range for searches of non-millisecond pulsars. However, recent pulse broadening
measurements \citep{spi13} and angular broadening measurements \citep{bow13} of the GC pulsar
have demonstrated that scattering may not be as severe a limitation.

The GC magnetar has the largest dispersion measure ever measured for a pulsar,
$1778 \pm 3$ cm$^{-3}$ pc, and the largest rotation measure ever measured for any object
other than Sgr A* itself, $-66960 \pm 50$ rad m$^{-2}$ \citep{eat13c,sha13}. \cite{eat13c}
have shown that the large Faraday rotation can be explained by a large magnetic field 
associated with the plasma within 10~pc of the GC black hole, suggesting a highly
complicated and magnetized interstellar medium in the GC region, well suited for scattering
electromagnetic radiation. \cite{spi13} measure a pulse broadening timescale of
$1.3 \pm 0.2$ s for J1745$-$2900 at 1 GHz, which, albeit large, is much less than predicted
values. The lack of detections of previous surveys implies either that previous surveys
have not been sensitive enough, or that the GC tends to produce magnetars. Given that observations of the GC have hitherto detected one magnetar and no normal
pulsar\footnote{In this paper, we use the term `normal pulsar' to mean a non-recycled
pulsar that is not a magnetar.}, in this paper, we attempt to constrain the number of
potentially observable pulsars (including magnetars) in the region using recent studies of the
pulsar luminosity function and spectral indices. We employ two complementary
methods\footnote{The software package that we developed to perform the analysis
described in this paper is available freely for download from\\
\url{http://psrpop.phys.wvu.edu/gcpulsars}.} --
firstly, a Bayesian parameter estimation approach, and secondly, a Monte Carlo (MC) approach to
constrain the pulsar population in the GC.

The organization of this paper is as follows. In \S\ref{sec_bayes}, we describe our Bayesian
technique and apply it to a few past surveys of the GC to obtain upper limits on the the number
of GC pulsars. In \S\ref{sec_mc}, we describe our MC approach and use it to constrain the number
of GC pulsars. In \S\ref{sec_disc}, we discuss our results and implications for future surveys.

\section{Constraining the GC pulsar content}

\subsection{Bayesian Approach}\label{sec_bayes}

To quantify the likely size of the pulsar population in the GC region, we treat the GC as
a population of pulsars at a common distance from the Earth, $D_{\rm GC}$, and consider
a survey of this region at some frequency $\nu$ as having some finite probability of
detecting a pulsar with flux density $S$ and a radio spectral index $\alpha$. Here, as
usual, we adopt a power-law relationship for the radio spectra \citep[see, e.g.,][]{lor95}
so that
$S \propto \nu^{\alpha}$. For a survey with some limiting sensitivity
$S_{\rm min}$ at frequency $\nu$, the corresponding limiting pulsar
pseudo-luminosity scaled to 1.4 GHz
\begin{equation}
L_{\rm min} = S_{\rm min} \left(\frac{1.4~{\rm GHz}}{\nu}\right)^{\alpha}
D_{\rm GC}^2.
\end{equation}
This limiting luminosity can be used to compute the detection probability, i.e., the
probability of observing a pulsar above this limit, based on a choice of the pulsar
luminosity function. \cite{fau06} have shown that the luminosity distribution of normal
pulsars in the Galactic field is log-normal in form. For the log-normal luminosity
function, the detection probability 
\begin{equation}
\theta = \frac{1}{2} \erfc*{\frac{\log{L_{\rm min}} - \mu}{\sqrt{2} \sigma}},
\end{equation}
where $\mu$ and $\sigma$ are the mean and standard deviation of the log-normal.

Using Bayes' theorem, we can utilize the above detection probability to estimate the number of
pulsars in the GC. The joint posterior probability density for the number of non-recycled
pulsars in the GC and the spectral index, given $n$ observed non-recycled pulsars,
\begin{equation}
p(N, \alpha | n) \propto p(n | N, \alpha)~p(N)~p(\alpha),
\end{equation}
where $p(n | N, \alpha)$ is the likelihood function, and $p(N)$ and $p(\alpha)$ are the
prior probability density functions of $N$ and $\alpha$, respectively. To account for
the fact that we have observed one magnetar in the GC and zero normal pulsars, the
observed number of non-recycled pulsars is written as $n = n_{\rm np} + n_{\rm mag}$,
where $n_{\rm np} = 0$ is the observed number of normal pulsars and $n_{\rm mag} = 1$ is
the observed number of magnetars. The aforementioned Bayesian relation can then be
rewritten, and the the likelihood function expanded, as
\begin{equation}
p(N, \alpha | n_{\rm np}, n_{\rm mag}) \propto p(n_{\rm np} | N, \alpha)
~p(n_{\rm mag} | N, \alpha)~p(N)~p(\alpha).
\end{equation}
Here we make the reasonable assumption of statistical independence for $n_{\rm np}$ and
$n_{\rm mag}$ given $N$ and $\alpha$, as the formation scenarios for normal pulsars and
magnetars are likely different.

The parameter that we are trying to constrain, namely, the total number of non-recycled
pulsars $N$, can be written in terms of the magnetar fraction $f$ -- the ratio of
magnetars to normal pulsars -- as
\begin{equation}
N = N_{\rm np} + N_{\rm mag} = N (1 - f) + N f.
\end{equation}
The value of $f$ is highly uncertain. In the Galactic field, 25 magnetars are
known\footnote{\url{http://www.physics.mcgill.ca/~pulsar/magnetar/main.html}}, at least
four of which emit in the radio, and of which only one was found in a radio survey
\citep{lev10}. There are about 2000 non-recycled pulsars in the
field\footnote{\url{http://www.atnf.csiro.au/people/pulsar/psrcat/}} \citep{man05},
giving $f \approx 0.01$. Considering only radio-emitting magnetars, $f$ becomes 0.002, and
considering only radio-loud magnetars detected in surveys, the magnetar fraction reduces to
0.0005. Due to the intermittency in the radio emission of magnetars, together with the
selection effects that plague radio surveys, the exact value of the magnetar fraction is
unknown. So we decided to parametrize $f$ in our analysis, giving
\begin{equation}
\begin{split}
p(N, f, \alpha | n_{\rm np}, n_{\rm mag}) \propto~&p(n_{\rm np} | N, f, \alpha)
~p(n_{\rm mag} | N, f, \alpha)\\
& \times p(N)~p(f)~p(\alpha).
\end{split}
\end{equation}

We compute the two likelihood functions using the binomial probability distribution,
following \cite{boy11}. In the case of normal pulsars, we have
\begin{equation}
p(n_{\rm np} | N, f, \alpha) = (1 - \theta)^{N (1 - f)},
\end{equation}
and for magnetars,
\begin{equation}
p(n_{\rm mag} | N, f, \alpha) = N f \gamma (1 - \gamma)^{N f - 1},
\end{equation}
where $\gamma$ is the magnetar detection probability. Magnetars are characterized by a
flat spectral index, i.e., their luminosities appear to be independent of the observing frequency.
Under this assumption we derive the magnetar detection probability from the normal pulsar detection
probability as
\begin{equation}
\gamma = \theta(\alpha = 0) = \frac{1}{2}
\erfc*{\frac{\log{(S_{\rm min} D_{\rm GC}^2)} - \mu}{\sqrt{2} \sigma}},
\end{equation}
and $p(n_{\rm mag} | N, f, \alpha)$ becomes $p(n_{\rm mag} | N, f)$.

To avoid any bias in our analysis, we adopt non-informative priors for $N$ and $f$ (i.e., uniform probability
within given ranges stated below). For $\alpha$, we
use the results of \cite{bat13} and take
\begin{equation}
p(\alpha) \propto e^{-\frac{(\alpha - \bar{\alpha})^2}{2 \sigma_{\alpha}^2}},
\end{equation}
where $\bar{\alpha} = -1.41$ is the mean spectral index and $\sigma_{\alpha} = 0.96$ is
the standard deviation. The final Bayesian relation can then be written as
\begin{equation}
\begin{split}
p(N, f, \alpha | n_{\rm np}, n_{\rm mag}) \propto~&(1 - \theta)^{N (1 - f)}~N f \gamma (1 - \gamma)^{N f - 1}\\
&\times~p(N)~p(f) ~e^{-\frac{(\alpha - \bar{\alpha})^2}{2 \sigma_{\alpha}^2}}.
\end{split}
\end{equation}
This is then integrated over $f$ and $\alpha$ to obtain the marginalized posterior of
$N$, $p(N | n_{\rm np}, n_{\rm mag})$.

\begin{figure}
\includegraphics[width=\linewidth]{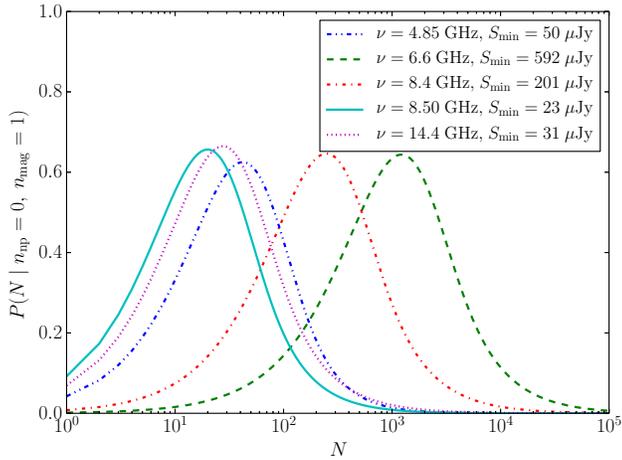}
\caption{Posterior probability density functions of $N$ for each of the surveys listed
in Table~\ref{tab_res}.
\label{fig_pdf}}
\end{figure}

We applied our technique to the surveys of the GC analyzed by \cite{wha12}, namely those
discussed in \cite{joh06}, \cite{den10}, \cite{mac10} and \cite{bat11}. In applying our
analysis to these past surveys, we made the reasonable assumption that, had the magnetar
become active earlier, all these surveys would have detected it. We computed the survey
sensitivity limits based on information provided in those papers, and additionally performed a
normalization to ensure that the minimum flux density values are average values over the part of
the beam that cover the inner 1 pc of the GC, modelling each beam as a gaussian. We used the
broad ranges of [1, 10$^5$] for $N$ and [0.001, 0.999] for $f$. We found that for the more
sensitive surveys \citep{den10,mac10}, the mean of the posterior on $N$ is in the range
800--3000 and the 99 per cent upper limit is in the range 12000--47000. Our results are
tabulated in Table~\ref{tab_res}. Figure~\ref{fig_pdf} shows the posterior probability density
functions of $N$ derived from each of the surveys listed in Table~\ref{tab_res}.

\begin{table}
\caption{Results of our analysis for a few GC surveys. $\left<N\right>$ is the expected
value of $N$ while $N_{\rm max}$ is the 99\% upper limit on $N$.
\label{tab_res}}
\begin{center}
\begin{tabular}{lllll}
\hline
$\nu$ & $S_{\rm min}$ & $\left<N\right>$ & $N_{\rm max}$ & Survey reference\\
(GHz) & ($\mu$Jy) & & &\\
\hline
\hphantom{0}4.85 & \hphantom{00}50 & \hphantom{00}800 & 12000 & \cite{den10}\\
\hphantom{0}6.6 & \hphantom{0}592 & 16000 & 92000 & \cite{bat11}\\
\hphantom{0}8.4 & \hphantom{0}201 & \hphantom{0}9000 & 86000 & \cite{joh06}\\
\hphantom{0}8.50 & \hphantom{00}23 & \hphantom{0}1200 & 21000 & \cite{den10}\\
14.4 & \hphantom{00}31 & \hphantom{0}3000 & 47000 & \cite{mac10}\\
\hline
\end{tabular}
\end{center}
\end{table}

\begin{figure}
\includegraphics[width=\linewidth]{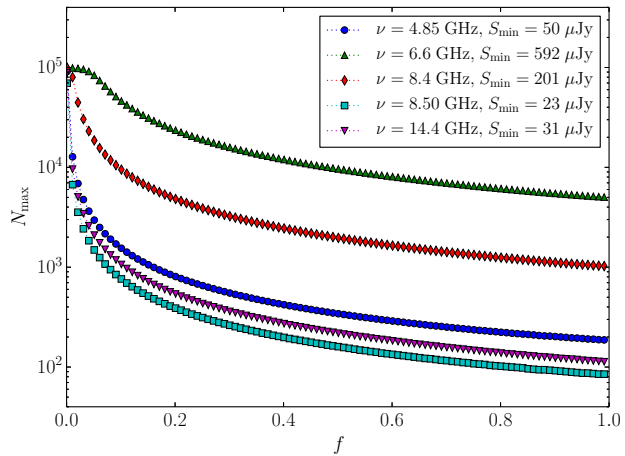}
\caption{The 99\% upper limit on $N$, $N_{\rm max}$, as a function of magnetar fraction,
$f$, for each of the surveys listed in Table~\ref{tab_res}.
\label{fig_nvsf}}
\end{figure}

As mentioned previously, the magnetar fraction in the Galactic field is uncertain, and that
in the GC is unknown. To study how the 99\% upper limit on $N$ would vary with magnetar fraction, instead of using a wide prior on $f$, we
chose delta functions in the range (0.0, 1.0), and for each of those magnetar fractions, we
computed the upper limit. Figure~\ref{fig_nvsf} shows the results of this analysis. For a small
magnetar fraction, the upper limits would be close to the values reported in
Table~\ref{tab_res}. If, on the other hand, formation of radio-loud magnetars are somehow
favoured in the GC region, the fact that we have detected one magnetar implies that we can
expect a smaller population size.

\subsection{Monte Carlo Approach}\label{sec_mc}

The Bayesian technique described above is relatively agnostic about the period and
luminosity evolution of normal pulsars. To make use of the results known from studies of
normal pulsars in the Galaxy \citep[see, e.g.,][]{fau06}, we perform an MC simulation of
the GC pulsar population and apply it to the  \cite{den10} survey. In this method, we
follow \cite{fau06} to simulate a population of $N_{\rm sim}$ pulsars, evolved over time
starting from the distributions of birth spin period, surface magnetic field at birth and
age. Picking $N_{\rm sim}$ from the range [$5 \times 10^3$, $8 \times 10^4$], we compute the number
of pulsars that have not crossed the death line, i.e., the number of radio-loud pulsars,
denoted by $N_{\rm GC}$. We then apply radiation beaming correction and compute the number
of potentially observable pulsars. Given that \cite{spi13} measure the scattering
timescale at 1~GHz as $\sim 1.3$ s, we scale it to the observation frequency with a power-law
spectral index $\alpha_\tau = -3.8$ to compute the scatter-broadened pulse widths for each
pulsar. We then compute the luminosities $L$ at 1.4 GHz, followed by computations of the
signal-to-noise ratio ($S/N$). We then apply the $S/N$ threshold of the survey to get the
detectable number of pulsars, denoted by $n_{\rm obs}$.

\begin{figure}
\includegraphics[width=\linewidth]{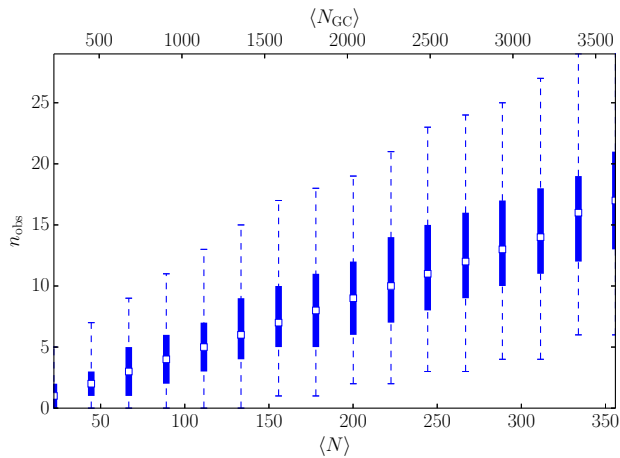}
\caption{The number of observed pulsars, $n_{\rm obs}$ versus the mean of the number of
potentially observable pulsars, $\left<N\right>$ and the mean of the total number of
radio-emitting pulsars in the GC, $\left<N_{\rm GC}\right>$. Each point represents 10$^4$ MC
realizations. The white markers indicate the mean of $n_{\rm obs}$ and the thick
error bars represent the corresponding 68.3 per cent confidence intervals. The dashed error
bars are 99.7 per cent confidence limits.
\label{fig_nvsN}}
\end{figure}

For each value of $N_{\rm sim}$, 10$^4$ MC realizations were generated to ensure stability
in the mean of $n_{\rm obs}$. Figure~\ref{fig_nvsN} shows a plot of observed number of
pulsars along with both 68.3 and 99.7 per cent confidence limits versus the mean value of
the number of potentially observable pulsars in the GC, $\left<N\right>$, and the mean value
of the total number of radio-emitting pulsars in the region, $\left<N_{\rm GC}\right>$. As
can be seen, the lower limits of the 99.7 per cent confidence intervals become inconsistent
with an actual detection of zero normal pulsars for $\left<N\right>$ around 150. Any number
$\left<N\right> \gtrsim 150$ would mean that more pulsars should have been observed. The
fact that none have been detected gives an upper bound to the number of potentially
observable pulsars in the GC. The assumptions that go into this simulation are that the
luminosity function of GC pulsars is the same as that of field pulsars, and that the age
distribution of pulsars in the GC follow the uniform distribution for field pulsars (i.e., a
constant formation rate). Although a burst of supernovae has been proposed to have occurred
in the GC $\sim$ 10 Myr ago \citep{sof94}, near-infrared observations have revealed some
evidence that the star formation rate in the region has been roughly constant over the past
$\sim$ 10 Gyr \citep{fig04}. If these assumptions, including those about the birth spin period
and magnetic field distributions are applicable to magnetars as well, the upper limit on the
potentially observable population size (of both normal pulsars and magnetars) increases to
approximately 200.

\section{Discussion}\label{sec_disc}

The analyses we have performed in this paper differ from that of \cite{wha12} mainly in
that we use a more realistic luminosity function. Whereas \cite{wha12} used a power-law
luminosity function, it is more appropriate to use the log-normal as found by
\cite{fau06}. We also take into account the recent discovery of one magnetar in the GC,
and, in the case of the Bayesian analysis, the fraction of magnetar to normal pulsars in
the field. An important assumption we make here is that scattering in the inner parsec is
uniform and is consistent with that of the line of sight to the GC magnetar. This appears to be
a reasonable assumption, given that \cite{bow13} have shown that the angular sizes of the
magnetar and Sgr A* are consistent with both sources being behind the same scattering screen. The
results of our Bayesian analysis suggest that the population of potentially observable pulsars in
the inner parsec of the GC could be as large as several thousand, whereas our MC analysis yields
an upper limit of $\sim$~200. The reason the Bayesian analysis yields a broader constraint is
because it makes fewer assumptions than the MC method. While the former only assumes a form for
the luminosity distribution, the latter makes assumptions about the spin-down behaviour and
formation rate of pulsars in the GC. We note that, for a typical radio pulsar beaming fraction
of $\sim$~10~per~cent \citep{tau98}, the total number of radio-emitting pulsars in the region, for
either method, would be an order of magnitude larger. As per the MC method, the value of
$\left<n_{\rm obs}\right>$ corresponding to an actual detection of one pulsar is $\sim$ 7. This is
consistent with the results of \cite{dex13} who use an estimate of the number of massive stars in
the GC, a model of natal kick velocity, and the observed interstellar scattering to get
$\left<n_{\rm obs}\right> \approx 10$. Our conservative upper limit of $\sim$ 200 suggests that
there may not be any detectable pulsar close enough to Sgr A* to probe the gravitational field of
the GC black hole. However, \cite{koc12} have suggested that pulsars in the inner parsec that are
not close enough to Sgr A* can still be useful in detecting intermediate and stellar mass black
holes in orbit around the GC black hole.

How many pulsars can we expect to see in future surveys of the GC? To answer this question,
we started with the most constraining posterior on $N$ obtained using the Bayesian method,
based on the \cite{den10} survey. We performed MC simulations similar to those described in
\S\ref{sec_mc}, with the following differences: (i) Instead of picking a set of equi-spaced
values, we picked $N$ randomly from the Bayesian posterior; (ii) We did not apply any beaming
correction, as the Bayesian posterior gives the number of potentially observable pulsars.
The rest of the simulation proceeds as described before, yielding an observed number of
pulsars $n_{\rm obs}$. We performed these MC simulations for a few hypothetical surveys
using the Green Bank Telescope (GBT), for each of its receivers from L-band to K-band. The
backend used was assumed to be able to sample the maximum instantaneous bandwidths supplied
by the receivers. For the GBT K-band Focal Plane Array receiver, we assumed that it was
configured to use the VEGAS\footnote{\url{http://www.gb.nrao.edu/vegas}} backend such that the
instantaneous bandwidth sampled is 8750 MHz. The sensitivity of each survey was calculated
based on a minimum signal-to-noise ratio of 8, a duty cycle of 10 per cent and an observation
time of 7 hours, which is approximately the duration for which the GC is visible from Green
Bank. The receiver temperature, receiver gain, number of polarizations and bandwidth used in our
sensitivity calculations were taken from the GBT Proposer's
Guide\footnote{\url{https://science.nrao.edu/facilities/gbt/proposing/GBTpg.pdf}}. The GC
background temperatures were calculated from peak flux densities and spectral indices reported in
\cite{law08}. For each survey, we repeated the simulation 10$^4$ times to ensure that the mean
value stabilizes, and computed the mean and standard deviation for the number of detections. Our
results are tabulated in Table~\ref{tab_predict}. Encouragingly, our results suggest that there
are some prospects for a detection with surveys of this sensitivity in the near future. We
caution, however, that in spite of the large number of pulsars we estimate to be present in the
GC, and the fact that these surveys have high sensitivity, we may yet detect no pulsar. 

As a self-consistency test, to verify that the Bayesian technique we developed in
\S\ref{sec_bayes} and used to predict $n_{\rm obs}$ as described above actually works, we
applied the above MC simulations to all the past surveys listed in Table~\ref{tab_res},
again using our most constraining posterior on $N$. We also applied it to the most
sensitive survey of the GC at 1.4 GHz, the Parkes Multi-beam Pulsar Survey
\citep[PMPS;][]{man01,mor02}. The results of these MC simulations are tabulated in
Table~\ref{tab_predict}. For all past surveys, we obtained 1-sigma limits of the number of
detected pulsars that are consistent with zero.

\begin{table}
\caption{Predictions for surveys, both past and future, based on our most constraining
posterior on $N$. Here, $n_{\rm obs}$ is the mean value of the number of detectable pulsars,
given along with 68.3 per cent confidence limits. Values have been rounded to the
nearest integer, and the lower limits have been truncated at 0.
\label{tab_predict}}
\begin{center}
\begin{tabular}{lllll}
\hline
Survey & $\nu$ & $S_{\rm min}$ & $T_{\rm GC}$ & $n_{\rm obs}$\\
& (GHz) & ($\mu$Jy) & (K) &\\
\hline
Past surveys\\
\hline
PMPS & \hphantom{0}1.374 & 3519 & 690 & $0 \pm 0$\\
\cite{den10} & \hphantom{0}4.85 & \hphantom{00}50 & 285 & $1^{+0}_{-1}$\\
\cite{bat11} & \hphantom{0}6.6 & \hphantom{0}592 & \hphantom{0}90 & $0 \pm 0$\\
\cite{joh06} & \hphantom{0}8.4 & \hphantom{0}201 & \hphantom{0}90 & $0 \pm 0$\\
\cite{den10} & \hphantom{0}8.50 & \hphantom{00}23 & 116 & $3^{+0}_{-3}$\\
\cite{mac10} & 14.4 & \hphantom{00}31 & 103 & $2^{+0}_{-2}$\\
\hline
Future GBT surveys\\
\hline
L-Band & \hphantom{0}1.45 & \hphantom{0}105 & 435 & $0 \pm 0$\\
S-Band & \hphantom{0}2.165 & \hphantom{00}75 & 373 & $1^{+0}_{-1}$\\
C-Band & \hphantom{0}5.0 & \hphantom{00}41 & 285 & $2^{+0}_{-2}$\\
X-Band & \hphantom{0}9.2 & \hphantom{00}17 & 116 & $3^{+0}_{-3}$\\
Ku-Band & 13.7 & \hphantom{00}14 & 103 & $4^{+0}_{-4}$\\
K-Band & 22.375 & \hphantom{00}10 & \hphantom{0}83 & $6^{+0}_{-6}$\\
\hline
\end{tabular}
\end{center}
\end{table}

Our MC technique described in \S\ref{sec_mc} yields a conservative upper limit of $\sim$ 200
potentially observable pulsars, whereas our Bayesian technique yields broader constraints that
are an order of magnitude larger. Further deep surveys of the GC in the radio, and monitoring
for X-ray outbursts from potential magnetars in the region will help conclusively establish the
size of the GC pulsar population.

\section*{Acknowledgements}

We would like to thank the referee, Simon Johnston, for constructive criticism and Ryan O'Leary and Ron Maddalena for useful discussions.

\end{document}